\documentclass[%
reprint,
amsmath,amssymb,
aps,
floatfix,
]{revtex4-2}

\usepackage{graphicx}
\usepackage{dcolumn}
\usepackage{bm}

\usepackage[separate-uncertainty=true]{siunitx}
\usepackage{physics}
\DeclareMathOperator{\sinc}{sinc}
\usepackage{float}
\usepackage{letltxmacro}
\LetLtxMacro{\oldcite}{\cite}
\renewcommand{\cite}[1]{\mbox{\oldcite{#1}}}
\usepackage{hyperref}
\usepackage{tabularx}
\begin{document}

\preprint{APS/123-QED}

\title{Cryogenic integrated spontaneous parametric down-conversion}

\author{Nina Amelie Lange$^{1,*}$, Jan Philipp Höpker$^{1}$, Raimund Ricken$^{2}$, Viktor Quiring$^{2}$,\\Christof Eigner$^{2}$, Christine Silberhorn$^{2}$, and Tim J. Bartley$^{1}$}
\affiliation{\vspace{1mm}$^1$\,Mesoscopic Quantum Optics, Department of Physics, Paderborn University,\\Warburger Str.~100, 33098 Paderborn, Germany}
\affiliation{\vspace{1mm}$^2$\,Integrated Quantum Optics, Department of Physics, Paderborn University,\\Warburger Str.~100, 33098 Paderborn, Germany}
\affiliation{\vspace{1mm}$^{*}$\,nina.amelie.lange@upb.de\vspace{1mm}}

\date{\today}

\begin{abstract}
Scalable quantum photonics relies on interfacing many optical components under mutually compatible operating conditions. To that end, we demonstrate that spontaneous parametric down-conversion (SPDC) in nonlinear waveguides, a standard technology for generating entangled photon pairs, squeezed states, and heralded single photons, is fully compatible with cryogenic operating conditions required for superconducting detectors. This is necessary for the proliferation of integrated quantum photonics in integration platforms exploiting quasi-phase-matched second-order nonlinear interactions. We investigate how cryogenic operation at \SI{4}{K} affects the SPDC process by comparing the heralding efficiency, second-order correlation function and spectral properties with operation at room temperature. 
\end{abstract}

\maketitle

Generation, manipulation, and detection of quantum light are based on a wide range of photonic quantum technologies~\cite{o2009photonic,wang2020integrated,moody2021roadmap}. In a fully-integrated platform, each technology must be mutually compatible~\cite{wang2020integrated}, not only with regard to the optical degrees of freedom of interest, but also their operating conditions. While nonlinear optics, in particular frequency conversion and electro-optic manipulation, is optimized for operation under ambient conditions, many quantum photonic technologies, in particular superconducting single-photon detectors and low-noise single emitters, require cryogenics~\cite{eisaman2011invited,natarajan2012superconducting}. To enable the mutual integration of these components, it is therefore necessary to adapt existing techniques and technologies to be functional in the same environment. 

One important and well-established technique for generating nonclassical light under ambient conditions is spontaneous parametric down-conversion (SPDC)~\cite{rarity1987observation,shih1988new,ou1988violation}. In various configurations, this can be used to generate heralded single photons~\cite{hong1986experimental}, entangled states~\cite{kwiat1995new}, and squeezed states~\cite{wu1986generation}. SPDC is a nonlinear interaction in a material exhibiting a second-order optical nonlinearity, in which a pump photon spontaneously decays into two daughter photons. Energy and momentum conservation (phase-matching) dictate the wavelengths at which these daughter photons are emitted. This process is well-established as a means to generate nonclassical light under ambient conditions. 

SPDC has been used extensively as a source of nonclassical light in materials such as lithium niobate, in which phase-matching is achieved through periodic inversion of the spontaneous polarization of the crystal~\cite{tanzilli2001highly,tanzilli2002ppln,fujii2007bright,krapick2016chip}. Furthermore, wave\-guides can be fabricated in this material, further enhancing the nonlinear interaction and allowing additional control of the phase-matching properties. Titanium in-diffused waveguides in particular have shown a wide range of functionality combining quantum light sources and electro-optic processing on low-loss chips~\cite{sohler2008integrated,sharapova2017toolbox,montaut2017high,luo2019nonlinear,thiele2020cryogenic}. Demonstrating this functionality  under cryogenic operating conditions is therefore vital to augment the range of integrated circuits that can be implemented.

In this paper, we show that cryogenic SPDC is indeed possible, and discuss variations to the phase-matching caused by the large temperature change. We show photon pair emission and joint spectral measurements when the sample is cooled to \SI{4.7}{K}. This is directly compared to the same sample at room temperature. The SPDC process we consider is quasi-phase-matched \mbox{Type-II} SPDC in titanium in-diffused waveguides in periodically poled lithium niobate~\cite{martin2010polarization,herrmann2013post,montaut2017high}. This interaction results in photons generated in orthogonal polarization modes, traditionally called signal and idler. In this configuration, the signal and idler modes can be deterministically separated, and the measurement of one photon can be used to herald the presence of the other. 

In order to understand SPDC under cryogenic conditions, we study the temperature-dependent variations in the spectral properties of signal and idler. These are dictated by energy and momentum conservation during the nonlinear interaction. Energy conservation is specified by the pump beam; it is independent of the sample temperature $T$. By contrast, momentum conservation is determined by the crystal length, its dispersion, and the poling period, all of which exhibit temperature dependence.

Momentum conservation is governed by the difference in propagation constants $k$ of the interacting modes, given by
\begin{equation}\label{eqn:PM}
\Delta k = k_p(\lambda_p) - k_s(\lambda_s) - k_i(\lambda_i) \,, 
\end{equation}
where $\lambda$ is the wavelength, and the subscripts $p,s,i$ denote the pump, signal and idler modes, respectively.

In general, due to the dispersion properties of the material, the interacting modes are not phase-matched, \mbox{{i.e.} $\Delta k\neq0$}. However, in many second-order materials, momentum conservation can be achieved by introducing an additional contribution to the momentum, arising from a periodic inversion of the crystal symmetry, known as periodic poling~\cite{tanzilli2002ppln}. Eq.~\ref{eqn:PM} is thus modified to ${\Delta k' = \Delta k \pm 2\pi/\Lambda}$, where $\Lambda$ is the poling period of the crystal. Thus, $\Lambda$ can be chosen such that $\Delta k'=0$ for a given combination of signal and idler wavelengths $\lambda_{s,i}$.

By writing $k=2\pi n/\lambda$, where $n$ is the refractive index, we can account for the temperature dependence of the interaction, as well as the thermal contraction of the poling period $\Lambda(T)$. Without loss of generality, the phase-matching condition may be written as
\begin{equation}\label{eqn:polingperiod}
\frac{n_{\mathrm{TE}}(\lambda_p,T)}{\lambda_p} - \frac{n_{\mathrm{TE}}(\lambda_s,T)}{\lambda_s} - \frac{n_{\mathrm{TM}}(\lambda_i,T)}{\lambda_i} - \frac{1}{\Lambda(T)} = 0 \,,
\end{equation}
where $n_{\mathrm{TE}}(\lambda,T)$ and $n_{\mathrm{TM}}(\lambda,T)$ are the temperature-dependent effective refractive indices of the transverse-electric (TE) and transverse-magnetic (TM) polarized modes.

For the following calculations, we use temperature dependent effective refractive index data based on Sellmeier equations for bulk lithium niobate~\cite{edwards1984temperature,jundt1997temperature}, which are extrapolated for low temperatures and modified for the waveguide geometry. An additional correction is applied to the extrapolated data, which is empirically obtained by measuring the second harmonic generation (SHG) spectra during the cool-down of the waveguide (for details on the extrapolation and measurement procedure, see~\cite{bartnick2021cryogenic}). Empirical data for the thermal expansion coefficient of lithium niobate in the $y$-direction is available for $T\geq\SI{60}{K}$~\cite{scott1989properties}. The expansion coefficient tends to zero at \SI{0}{K}, therefore when extra\-polating for temperatures below \SI{60}{K}, the sample length and the poling period can be assumed constant.

\begin{figure}[htbp]
\centering\includegraphics[width=0.92\linewidth]{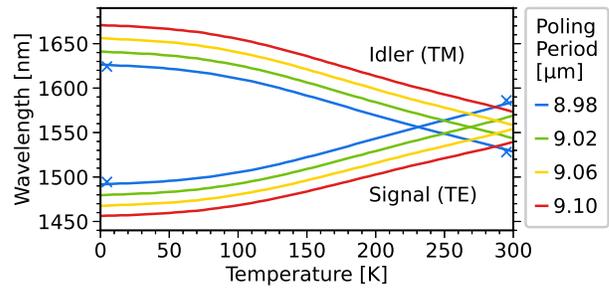}
\caption{\label{fig:pdcwavelengthshift}Theoretical temperature dependence of signal and idler wavelengths for a pump wavelength of $\lambda_p = \SI{778}{nm}$, displayed for different poling periods. The crosses show our spectral measurement results for a poling period of \SI{8.98}{\micro m}.}
\end{figure}

We calculate the changes in the phase-matching properties for temperatures in a range from \SI{300}{K} down to \SI{0}{K} by computing the signal and idler wavelength pair $(\lambda_s,\lambda_i)$ which fulfills energy conservation and temperature dependent momentum conservation (Eq.~\ref{eqn:polingperiod}). The calculation is performed for a pump wavelength of ${\lambda_p=\SI{778}{nm}}$ and for four poling periods, which are chosen to enable spectrally degenerate SPDC around room temperature (see Fig.~\ref{fig:pdcwavelengthshift}). From room temperature to \SI{4}{K}, the wavelengths of signal and idler are expected to shift by about \SI{90}{nm}. 

In fact, the temperature dependence of phase-matching is often exploited to achieve a desired nonlinear interaction, by placing the sample on a temperature controlled mount~\cite{montaut2017high}. Under cryogenic operation, thermal tuning is not possible, therefore Fig.~\ref{fig:pdcwavelengthshift} shows the importance of achieving the correct poling period for operation at a fixed temperature.

These calculations account for macroscopic, well-defined changes to the refractive indices. In many second-order materials, other temperature-dependent effects may change the resulting spectral behavior. In the case of lithium niobate, the material is pyroelectric, piezoelectric, and photorefractive~\cite{weis1985lithium,volk2008lithium}, meaning temperature changes (rather than absolute temperature) and optical power can locally alter the refractive index and therefore the phase-matching. Previous work has shown this may be a transient effect during temperature cycling, which nevertheless exhibits stable operation under constant temperature~\cite{bartnick2021cryogenic}. 

Building on these calculations, we experimentally investigated the phase-matching and other source properties of a lithium niobate wave\-guide chip operated under cryogenic conditions. The wave\-guide sample is fabricated by titanium in-diffusion into a \mbox{$z$-cut} congruently-grown uncoated lithium niobate chip. These wave\-guides support low-loss propagation of TE and TM polarization modes, with losses down to \SI{0.03}{dB/cm} at \SI{1550}{nm} under room temperature operation~\cite{sohler2008integrated,hopker2021integrated}. A single chip of length \SI{24.4}{mm} contains 16 groups of three straight waveguides, and each waveguide group is periodically poled with poling periods from \SI{8.98}{\micro m} to \SI{9.12}{\micro m} in increments of \SI{0.02}{\micro m}. At room temperature, these poling periods allow for degenerate \mbox{Type-II} SPDC with the signal and idler wavelength in the telecom \mbox{C-band}. 

\begin{figure*}[hbtp]
\centering\includegraphics[width=0.97\textwidth]{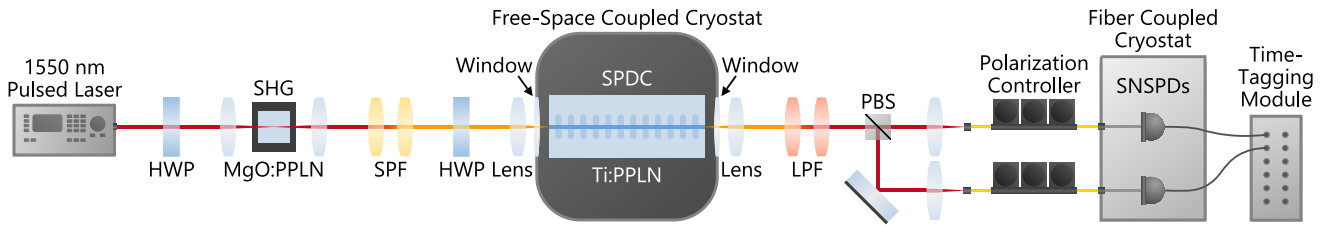}
\caption{\label{fig:setupPDC}Experimental setup to measure SPDC photons, generated in the periodically poled lithium niobate waveguide, which is positioned inside a free-space coupled cryostat. HWP: half-wave plate, MgO:PPLN/Ti:PPLN: MgO-doped/titanium-indiffused periodically poled lithium niobate, SPF/LPF: shortpass/longpass filter, PBS: polarizing beam splitter.}
\end{figure*}

The experimental setup is shown in Fig.~\ref{fig:setupPDC}. We place our waveguide sample inside a free-space coupled cryostat, with a base temperature of \SI{4.7}{K}. The waveguide end facets are optically accessible through transparent windows which allows us to couple the laser beam to the waveguide with aspheric lenses positioned outside the cryostat. The in- and out-coupling lenses are anti-reflection coated for the pump and down-converted light, respectively. We can move the sample inside the cryostat with a motorized mount to change between waveguides and therefore different poling periods at any time.

In order to prepare the pump beam for the SPDC process, we use the SHG signal from a bulk periodically poled MgO-doped lithium niobate (MgO:PPLN) crystal (heated to about 70~$^\circ$C). We pump the SHG with an ultrashort pulsed infrared laser with a repetition rate of \SI{80}{MHz}. The pump spectrum exhibits a Gaussian shape with a central wavelength of \SI{778.0(01)}{nm} and a bandwidth of \SI{3.2(01)}{nm}.

We set the polarization of the pump light to excite the TE waveguide mode, in order to pump Type-II SPDC. Following the chip, long-pass filters (LPF) remove the high-energy pump beam, and the signal and idler photons are separated via a broadband polarizing beam splitter (PBS) before being coupled into single-mode fibers. We use superconducting nanowire single-photon detectors (SNSPDs) located in a separate cryostat to measure the photons. Single counts and coincidences of signal and idler are recorded with a time-tagging module. 

We characterize our SPDC source at room temperature (\SI{295}{K}) and under cryogenic conditions (\SI{4.7}{K}). The free-space coupled cryostat shown in Fig.~\ref{fig:setupPDC} allows us to use exactly the same setup for both measurements. Moreover, it enables us to optimize the beam coupling to the waveguide end facets at any time. We characterize our source with regard to the spectral properties and the source performance metrics.

\begin{figure}[!bh] 
\centering\includegraphics[width=0.96\linewidth]{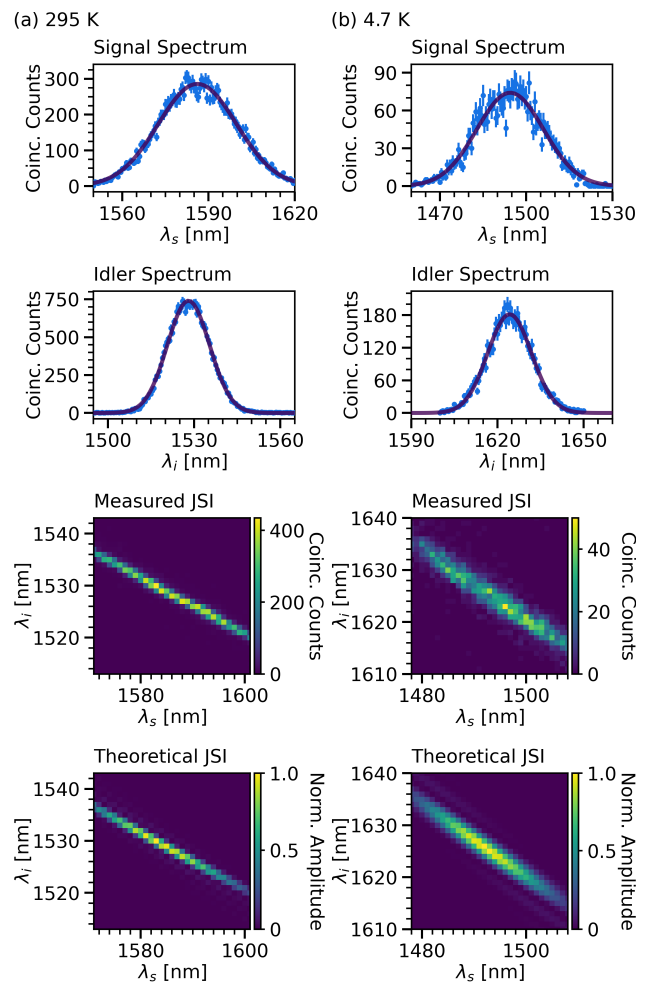}
\caption{\label{fig:spectralResults}Experimental results for the spectral measurements of the signal and idler photons, together with the simulated joint spectral intensity (JSI) for a room temperature poling period of \SI{8.98}{\micro m} (for higher resolution, see Fig.~\ref{fig:JSISimulation}). The marginal spectra and the JSI are measured at room temperature (a), and under cryogenic conditions (b). The solid line in each marginal spectra represents a Gaussian fit to the measurement data and the errorbars correspond to Poisson errors.} 
\end{figure}
We investigate the spectral properties by employing a home-built scanning-grating spectrometer setup (for details, see Appendix~\ref{sec:appA}). We measure the marginal spectra by inserting one spectrometer into the signal or idler path in front of the polarization controller. Afterwards, we perform a measurement of the joint spectral intensity (JSI) by applying one spectrometer to the signal path, and another one to the idler path. These results are shown in Fig.~\ref{fig:spectralResults}. 

Our results show that cryogenic operation of SPDC is possible and that the spectral properties behave as expected. For both measurements, the marginal spectra exhibit a Gaussian shape and the JSI is represented by an elongated ellipse. The obtained central wavelengths of signal and idler are summarized in Table~\ref{tab:results}. The measured wavelengths are in very good agreement with our simulation of a wavelength shift of about \SI{90}{nm} (compare Fig.~\ref{fig:pdcwavelengthshift}). 

\begin{table*}[t]
\centering
\caption{\label{tab:results}Comparison of our source performance at room temperature (295~K) and under cryogenic conditions (4.7~K). The table shows the phase-matched signal and idler wavelengths ($\lambda_s$, $\lambda_i$), the effective crystal length $L_{eff}$, the brightness $B$, the Klyshko efficiency $\eta_{\mathrm{Klyshko}}$, the coincidences-to-accidentals ratio CAR, and the heralded autocorrelation function $g_h^{(2)}(0)$.}
\renewcommand{\arraystretch}{1.5}
\begin{tabularx}{1.0\textwidth}{c|ccccccc}
	\hline
	& $\lambda_s$~[nm] & $\lambda_i$~[nm] & $L_{eff}$~[mm] & Brightness $B~\big[\frac{\mathrm{pairs}}{\mathrm{s~mW}}\big]$ & $\eta_{\mathrm{Klyshko}}~[\%]$ & CAR & $g_{h}^{(2)}(0)$ \\
	\hline
	295~K & $1586.14\pm0.12$ & $1528.05\pm0.04$ & $7.3\pm0.3$ & $(6.6\pm0.3)\times10^5$ & $8.47\pm0.03$ & $53.9\pm0.2$ & $0.033\pm0.004$ \\
	4.7~K & $1494.45\pm0.19$ & $1624.24\pm0.09$ & $3.65\pm0.05$ & $(3.3\pm0.1)\times10^5$ & $4.77\pm0.03$ & $32.9\pm0.2$ & $0.07\pm0.01$    \\
	\hline
\end{tabularx}
\end{table*}
\renewcommand{\arraystretch}{1}
The Gaussian fits provide the spectral bandwidths $\Delta \lambda_{s}$ and $\Delta \lambda_{i}$ of signal and idler. The signal bandwidth decreases during the cooldown from $\Delta \lambda_s = \SI{32.08(028)}{nm}$ to ${\Delta \lambda_s = \SI{27.4(05)}{nm}}$, while the cryogenic idler bandwidth ${\Delta \lambda_i = \SI{17.94(021)}{nm}}$ is unchanged compared to the room temperature result of ${\Delta \lambda_i = \SI{17.27(010)}{nm}}$. The decrease in the signal bandwidth matches the slight change in the angle of the measured JSI, which agrees with our theoretical predictions. In order to simulate the JSI, we take into account the pump wavelength and spectral width, the poling period, and the effective refractive indices. We keep the effective length of the waveguide as a variable parameter and perform an optimization until the simulation fits the measured JSI best (for details on the simulation, see Appendix~\ref{sec:appB}). According to the optimization, the effective length decreases from \SI{7.3(03)}{mm} to \SI{3.65(005)}{mm} when cooling down the sample. We expect that the decrease in effective length is due to photorefractive, pyroelectric, and piezoelectric effects, which can impair the mode guiding properties of the waveguide.

In addition to the spectral features of the cryogenic source, we compare the source performance metrics at both room temperature and \SI{4.7}{K}. We study the brightness $B$, the Klyshko (heralding) efficiency $\eta_\mathrm{Klyshko} $, the coincidences-to-accidentals ratio CAR, and the heralded autocorrelation function $g_h^{(2)}(0)$. These results are summarized in Table~\ref{tab:results}.

The brightness of our source is given by ${B = C_{si}/P_\mathrm{trans}}$, where $C_{si}$ is the coincidence rate of signal and idler, and $P_\mathrm{trans}$ is the transmitted pump power. The brightness of a wave\-guide source scales with the effective length of the sample~\cite{meyer2018high}. Our spectral measurements indicate a reduction of the effective length by approximately half, which indeed results in a commensurate reduction in the brightness by the same factor. 

The reduced brightness also affects the signal-to-noise of the source, which is evident in the Klyshko efficiency ${\eta_\mathrm{Klyshko} = \sqrt{C_{si}^2/(C_s C_i)}}$, where $C_s$ and $C_i$ are the single count rates. Our results show the Klyshko efficiency at cryogenic temperatures is roughly halved compared to room temperature, consistent with the reduced brightness at constant noise. 

A decrease in the signal-to-noise ratio is further verified by investigating the coincidences-to-accidentals ratio, which, in the low generation probability regime, is given by ${\mathrm{CAR} = (C_{si} R_\mathrm{rep})/(C_s C_i)}$, where $R_\mathrm{rep}$ is the laser repetition rate. Compared to room temperature, we observe a lower CAR value at \SI{4.7}{K} by a factor of approximately $\sqrt{2}$, which is consistent with the reduced effective length.

Finally, we measured the heralded autocorrelation function to investigate the photon number purity of our source. We add a 50:50 fiber beam splitter to the signal path in front of the polarization controller. For this configuration, the heralded autocorrelation function is calculated by ${g_h^{(2)}(0) = (C_{s_1s_2i} C_i)/(C_{s_1i} C_{s_2i})}$, where $C_{s_1i}$ and $C_{s_2i}$ are the coincidence rates of the two signal photons with the idler photons, and $C_{s_1s_2i}$ are the threefold coincidences. The heralded $g_h^{(2)}(0)$ remains well below the classical threshold of 1, but increases by a factor of two with respect to the room temperature value.

At cryogenic temperatures, all figures of merit are consistent with a reduction in the signal-to-noise ratio by a factor of two, compared with room temperature operation. We expect this decrease to be due to photorefractive effects which distort the guided mode~\cite{rams2000optical}, depending on laser intensity and exposure time. This distortion reduces conversion efficiency. Nevertheless, the figures of merit demonstrate a high-quality SPDC source at cryogenic temperature. 

Demonstrating mutual compatibility of operating conditions is crucial for the proliferation of quantum technologies. As part of this process, we demonstrated that spontaneous parametric down-conversion in quasi-phase-matched waveguides is compatible with the operating temperatures required for superconducting detectors. Despite changing the operating temperature by nearly two orders of magnitude, the source remained fully operational. This makes our source competitive for a wide variety of fully-integrated quantum circuits.

\begin{acknowledgments}
This project is supported by the Bundesministerium für Bildung und Forschung (BMBF), Grant Number 13N14911.
\end{acknowledgments}

\appendix

\section{Scanning-grating spectrometer design}\label{sec:appA}
\begin{figure}[htbp]
\centering\includegraphics[width=0.87\linewidth]{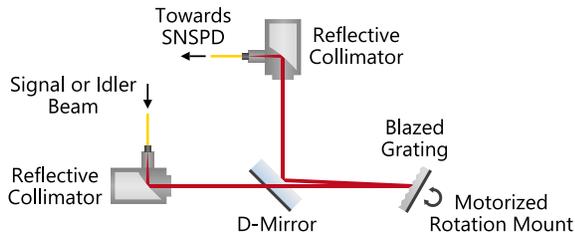}
\caption{\label{fig:setupSpectrometer} Design of the scanning-grating spectrometer setup, used to resolve the marginal spectra of signal and idler and the joint spectral intensity.}
\end{figure}

We investigate the spectral properties of our spontaneous parametric down-conversion (SPDC) source by employing a home-built scanning-grating spectrometer with single-photon resolution. The setup of our spectrometer is shown in Fig.~\ref{fig:setupSpectrometer}. The design comprises a diffraction grating, placed in a motorized rotation mount, which allows us to scan the incident angle. We filter the back-reflected wavelength components by coupling to a single-mode fiber.
To enable wavelength-insensitive fiber coupling, we use reflective collimators at the input and output, which are based on an off-axis parabolic mirror. This collimator design enables wavelength-independent collimation, which is important to achieve comparable performance given the expected wavelength shifts of up to \SI{100}{nm} across the temperature range. The fiber-to-fiber throughput of this device is \SI{66(2)}{\%}, and it exhibits a transmission bandwidth of \SI{0.909(0007)}{nm}, which was verified independently using a tunable diode laser at the wavelength range of interest (\SI{1440}{nm}-\SI{1640}{nm}). We achieve broadband spectral resolution at the single-photon level with extremely low noise by counting the transmitted photons with a superconducting nanowire single-photon detector (SNSPD).

We measure the marginal spectra by applying one spectrometer to the signal or idler path, respectively. The other path remains unfiltered. In order to keep uncorrelated noise counts as low as possible, we detect a heralded spectrum by measuring the transmitted photons in coincidence with the herald photons. Thus, the coincidence counts drop to approximately zero at the edge of the marginal spectra (see Fig.~\ref{fig:spectralResults}).

To measure the joint spectral intensity, we insert one spectrometer into the signal path and another one into the idler path. We scan the signal wavelength over a range of \SI{30}{nm} with a resolution of \SI{1}{nm}, while performing a full measurement of the idler spectrum for every signal wavelength step. This way, by measuring the coincidence counts of signal and idler, we capture the JSI of our SPDC source.

\section{Theoretical simulations}\label{sec:appB}
In order to evaluate our spectral measurement results and to estimate the effective length of our waveguide, we perform a simulation of the joint spectral intensity (JSI) at room temperature and at \SI{4.7}{K}.

We start with the description of the photon-pair state generated by an SPDC source, which can be written as~\cite{grice2001eliminating}
\begin{equation}\label{eqn:pdcstate}
\ket{\psi}_{\text{PDC}} \propto \iint \text{d}\omega_s\text{d}\omega_i \, f(\omega_s,\omega_i) \, \hat{a}_s^\dagger(\omega_s) \hat{a}_i^\dagger(\omega_i) \ket{0} \,,
\end{equation}
where $\hat{a}_s^\dagger(\omega_{s})$ and $\hat{a}_i^\dagger(\omega_{i})$ are the photon creation operators of the frequency modes $\omega_{s}$ and $\omega_{i}$. The term $f(\omega_s,\omega_i)$ is the joint spectral amplitude (JSA), which provides a full description of the spectral properties of signal and idler. It combines the pump distribution function $\alpha(\omega_s + \omega_i)$ and the phase-matching function $\Phi(\omega_s,\omega_i)$, associated with the energy and momentum conservation of the nonlinear interaction. The JSA is defined by
\begin{equation}\label{eqn:jsa}
f(\omega_s,\omega_i) = N\alpha(\omega_s + \omega_i) \Phi(\omega_s,\omega_i) \,,
\end{equation}
where $N$ is a normalization constant.

Since our pump spectrum exhibits a Gaussian shape, we write the pump distribution function, for a pair of signal and idler frequencies $\omega_s$ and $\omega_i$, as
\begin{equation}\label{eqn:alpha}
\alpha(\omega_s + \omega_i) = \exp \left[ -\dfrac{((\omega_s+\omega_i)-\omega_p)^2}{2 \sigma^2} \right] \,,
\end{equation}
where $\omega_p$ is the central pump frequency, and $\sigma$ is the standard deviation, directly related to the spectral bandwidth $\Delta\omega_p$, according to $\Delta\omega_p = 2\sqrt{2\ln{2}} \sigma$. We identify $\omega_p$ and $\Delta\omega_p$ by measuring the intensity profile of our pump beam with a commercially available optical spectrum analyzer.

The phase-matching function is given by~\cite{migdall2013single}
\begin{equation}\label{eqn:phi}
\Phi(\omega_s,\omega_i) = \sinc \left[ \Delta k'(\omega_s,\omega_i) \, \frac{L}{2} \right] \,,
\end{equation}
where $L$ is the effective length of the waveguide, and $\Delta k'$ is the phase-mismatch of the propagation constants. By applying Eq.~\ref{eqn:PM} together with the modification ${\Delta k' = \Delta k - 2\pi/\Lambda}$ and writing the wavelength as ${\lambda = c/(2\pi\omega})$, we can express the phase-mismatch by

\begin{eqnarray}\label{eqn:mismatch}
\Delta k'(\omega_s,\omega_i) =&& \frac{n_{\mathrm{TE}}(\omega_s + \omega_i) \cdot (\omega_s + \omega_i)}{c}\\
&-& \frac{n_{\mathrm{TE}}(\omega_s) \cdot \omega_s}{c} - \frac{n_{\mathrm{TM}}(\omega_i) \cdot \omega_i}{c} - \frac{2\pi}{\Lambda}\,.\nonumber
\end{eqnarray}

While the pump distribution function is independent of temperature, we need to take the temperature dependence of the phase-matching function into account. Therefore, we include the thermal contraction of the interaction length $L(T)$ and the poling period $\Lambda(T)$, as well as the temperature dependence of the effective refractive indices $n(\omega,T)$, into our calculation of Eq.~\ref{eqn:phi}.

We perform the simulation of the JSI for a two-dimensional array of frequencies ($\omega_s$,$\omega_i$), centered around $\omega_p/2$. For every frequency pair ($\omega_s$,$\omega_i$), we calculate the pump distribution function for our pump laser source from Eq.~\ref{eqn:alpha}, and the phase-matching function for our waveguide sample from Eq.~\ref{eqn:phi}. Next, we multiply the amplitudes of $\alpha(\omega_s + \omega_i)$ and $\Phi(\omega_s,\omega_i)$, according to Eq.~\ref{eqn:jsa}. The JSI is then obtained from the JSA by taking the absolute square: $\mathrm{JSI} = |f(\omega_s,\omega_i)|^2$.

\begin{figure*}[htpb] 
\centering\includegraphics[width=0.75\linewidth]{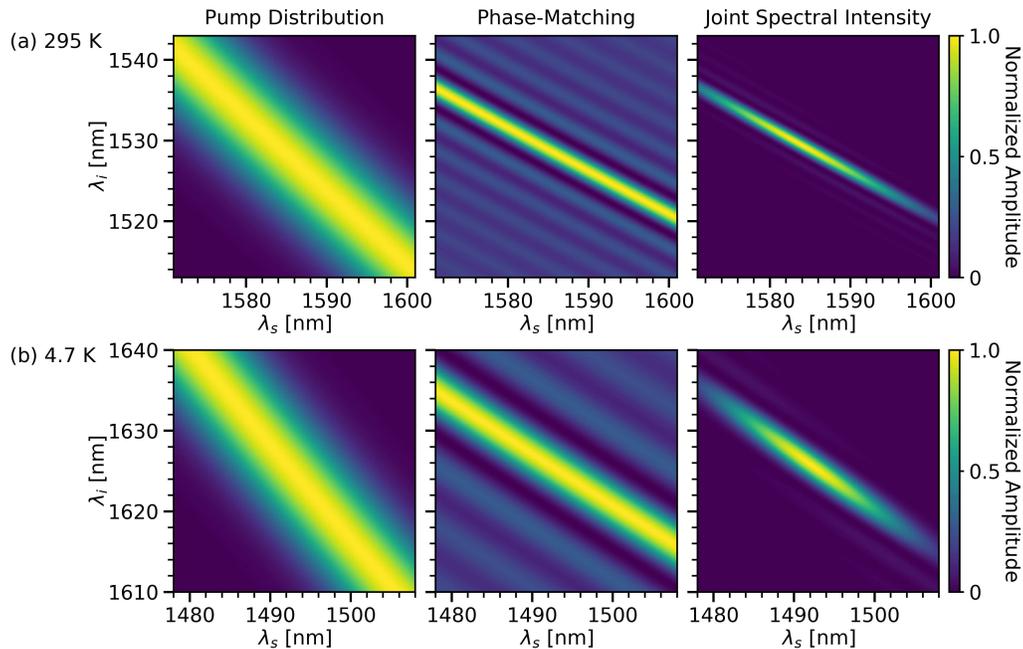}
\caption{\label{fig:JSISimulation}Optimized simulations of the joint spectral intensity, combining the pump distribution function and the phase-matching function, for operating the SPDC source at room temperature (a) and under cryogenic conditions (b).} 
\end{figure*}
In order to determine the effective length of the wave\-guide, we keep $L(T)$ in Eq.~\ref{eqn:phi} as a variable parameter, with an upper limit set to the length of the complete waveguide. While keeping all other parameters fixed, we simulate the JSI with the same resolution as our measurement data for different effective lengths. We perform an optimization by comparing the simulated JSIs to the normalized measured JSI and determining the simulation with the smallest standard deviation. The resulting joint spectral intensities, for a sample temperature of \SI{295}{K} and \SI{4.7}{K}, are shown in Fig.~\ref{fig:spectralResults}. In Fig.~\ref{fig:JSISimulation} we display the same simulated JSIs, together with the corresponding pump distribution and phase-matching functions, in a higher resolution.

Based on the simulations that best matched our measured data, we assume a decrease in the effective length from \SI{7.3(03)}{mm} to \SI{3.65(005)}{mm}. It can be seen from Fig.~\ref{fig:JSISimulation} that the decreased effective length at cryogenic temperatures corresponds to a broadening of the phase-matching function, and therefore the JSI. This broadening is also clearly visible in our measured joint spectral intensities shown in Fig.~\ref{fig:spectralResults}. Moreover, the simulations verify the slight change in the angle of the measured JSI, since there is also a change in the slope of the pump distribution function, when displayed in dependence of signal and idler wavelengths. Compared to the pump distribution function, there is only a very small change in the angle of the phase-matching function, which results from a change in the group velocities of the interacting modes.

\bibliography{library}

\end{document}